\documentclass[epjCONF]{svjour}
\usepackage{graphics}
\usepackage[varg]{txfonts} 
\usepackage[latin1]{inputenc}
\usepackage[usenames]{color}
\session-title{Conference Title, to be filled}
\begin{document}
\title{Rendez-vous of dwarfs}
\author{R.I.~Uklein\inst{1}\fnmsep\thanks{\email{uklein@sao.ru}} \and D.I.~ Makarov\inst{1} \and S.~Roychowdhury\inst{2} }
\institute{Special Astrophysical Observatory of Russian Academy of Sciences \and National Centre for Radio Astrophysics, Tata Institute of Fundamental Research}
\abstract{
We present observations of multiple system of dwarf galaxies at the Russian 6-m telescope and the GMRT (Giant Metrewave Radio Telescope). 
The optical observations are a part of the programme Study of Groups of Dwarf Galaxies in the Local Supercluster.
The group of galaxies under consideration looks like filament of 5 dwarfs.
Two faint galaxies show peculiar structure.
Long slit spectrum reveals inner motions about 150 km/s in one of them.
It suggests that the galaxy is on stage of ongoing interaction.
Probably, we see the group in moment of its formation.
} 
\maketitle
\section{The main parameters of the group}
\label{intro}
During the work on the catalog of groups of galaxies Makarov et al. (2010) \cite{makarov} have been found interesting groups of dwarf galaxies. One of these groups we have studied in detail at the 6-m telescope and GMRT.

\begin{figure}    
\begin{center}
    \resizebox{0.6\columnwidth}{!}
    {\includegraphics{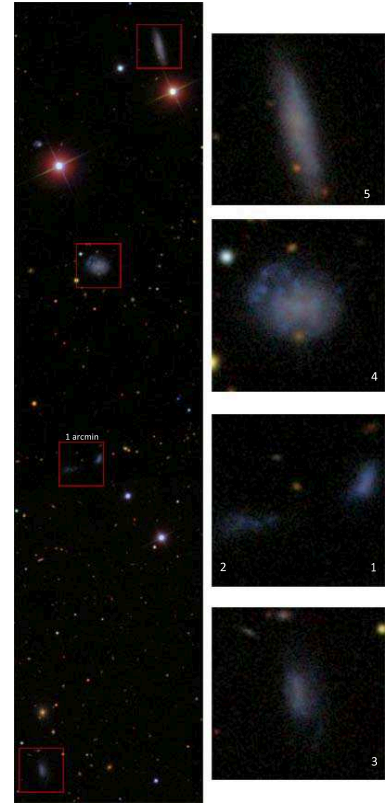} }
    \caption{
A SDSS DR7 image of the studied galaxy group. Left: the image of
the entire group. The red squares mark the regions with angular
sizes of 1 arcmin. Right: the same regions magnified with $\times$4. The
number of the galaxy in the group is indicated in the plots.
}\label{g33image}
\end{center}
\end{figure}

The SDSS image of the group is presented in Fig.~\ref{g33image}. 
The parameters of individual galaxies are indicated in Table~\ref{tab:1}.
The columns contain the following data: the number in the group, the
name of the galaxy (for SDSS galaxies the coordinates in the name are
omitted), the coordinates at the epoch J2000.0, mean radial velocity
of the group relative to the centroid of the group, total magnitude in
the SDSS \textit{g}-band, absolute \textit{g}-band magnitude, and luminosity in units of $10^8 L_\odot$.
\begin{table}
\caption{List of galaxies.}
\label{tab:1}       
\begin{tabular}{ cccccc }
\hline\noalign{\smallskip}
No.& RaDec(J2000) & $V_{LG}$, & g, & $M_{abs}$, & $L_g$, \\
   &  & km/s &  mag & mag & $10^8L_\odot$ \\
\noalign{\smallskip}\hline\noalign{\smallskip}
1 & 124412.1+621019 & 2682 & 17.78 & -15.15 & 1.28 \\
2 & 124418.0+621007 & 2650 & 18.08 & -14.82 & 0.95 \\
3 & 124423.2+620306 & 2660 & 17.53 & -15.38 & 1.58 \\
4 & 124412.0+621451 & 2614 & 15.82 & -17.05 & 7.38 \\
5 & 124359.9+621960 & 2602 & 16.16 & -16.70 & 5.35 \\
\noalign{\smallskip}\hline
\end{tabular}
\end{table}

The projection distance between the outermost galaxies,
determining the size of the group amounts to 190 kpc. The mean
velocity of the group, weighted over the velocity errors, amounts
to 2675 km/s. Total luminosity of all galaxies in the SDSS
\textit{g}-band is approximately  $1.6\cdot 10^9 L_\odot$.
The velocity dispersion for this group is approximately 20
km/s. Total mass, calculated via the projected mass estimator
method, is equal to  $1.5\cdot 10^{11} M_\odot$ and the $M/L$ ratio is about 110 in solar units in \textit{g}-band.

The groups of our sample have small dispersions of velocity and projected distances between galaxies. It makes them look like the assosiations of dwarf galaxies from Tully et al. (2006) \cite{tully}.

 \section{Observations}
  \label{Astr_telesc}
\subsection{BTA}

\begin{figure}
\begin{center}
    \resizebox{0.9\columnwidth}{!}{%
    \includegraphics{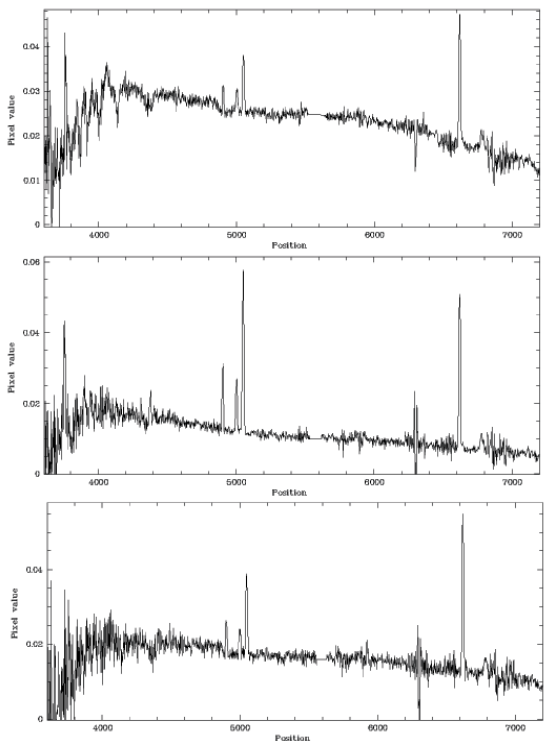}      }
    \caption{
BTA/SCORPIO spectra for the 1st and 2nd galaxies at the wavelength
range of $3600-7200$~\AA{}. The upper plot represents the
spectrum of the first galaxy, the middle and bottom plots --- the
second galaxy. The spectra of the second galaxy have a velocity
difference of about 150 km/s, what indicates the interaction effects
in this galaxy.}
\label{spectra_fig}
\end{center}
\end{figure}

We observed 1st and 2nd galaxies on the 6-m BTA telescope on 
August, 2009 with the SCORPIO focal reducer \cite{scorpio}. We used the
VPHG550G grism and a long slit with the size of $1^{\prime\prime} \times 360^{\prime\prime}$. The
wavelength range of the grism is $3100-7300$~\AA{}. The long slit was simultaneously centered on
the brightest regions of both galaxies. After the standard
processing of two-dimensional spectra in the ESO-MIDAS package we extracted 3 one-dimensional spectra. The spectra are shown in
Fig.~\ref{spectra_fig}. 
In general the spectra are emission dominated but the
spectrum of 1st galaxy has appreciable absorptions in the hydrogen
lines. Apart from emissions in the hydrogen lines, the spectra
contain the emission lines of [OIII], [SII] and others. We
determined the metallicity (oxygen abundance) with a reasonable
accuracy for only one spectrum (Fig.~\ref{spectra_fig}, middle panel). The value
of 12+log(O/H) is 7.2 $\pm{}$ 0.1. To find the metallicity we use a
direct method based on [OIII] $\lambda 4363/(\lambda 4959+\lambda 5007)$ ratio \cite{izotov}.

Hence, it is a rather low-metallicity galaxy (a little higher metallicity than
that of I Zw 18). Spectrum of 1st galaxy shows sign of low metallicity too.
An important feature of the 2nd galaxy is a difference between the radial velocities of parts of this galaxy. It is amounts to 150 km/s.
\vspace{1cm}

\subsection{GMRT}
\begin{figure}    
\begin{center}
    \resizebox{1.0\columnwidth}{!}{%
    \includegraphics{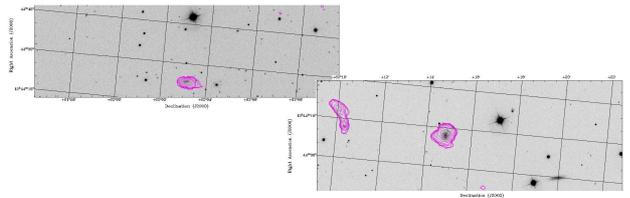}      }
    \caption{
GMRT HI map of the group with resolution 10k$\lambda$. The beam of the HI map is 10k$\lambda$: 25.11 x 22.18 arcsec, PA = $-$11.30$^\circ $.}
\label{gmrt_fig}
\end{center}
\end{figure}

Fig.~\ref{gmrt_fig} present GMRT data obtained on April 2010.

Three clouds of HI were observed in the field, one covering galaxies numbered 1 and 2, one covering galaxy no 3, and another covering galaxy no 4. Basic parameters for these three clouds presented in Table~\ref{tab:2}.

\begin{table}
\vspace{-0.5cm}
\caption{Total HI masses.}
\label{tab:2}       
\begin{tabular}{ccc}
\hline\noalign{\smallskip}
Galaxy & Velocity spread, km/s & HI mass,$10^8 M_\odot$ \\
\noalign{\smallskip}\hline\noalign{\smallskip}
1, 2 & 2460-2543  & 2.15  \\
3   & 2361-2523  & 4.81  \\                                                
4   & 2369-2488  & 3.74  \\                                                
\noalign{\smallskip}\hline
\end{tabular}
\end{table}
\vspace{-0.1cm}
  \section{Conclusions}
The mass-to-luminosity ratio for the group is more than 100 $M_\odot/L_\odot$. It is  indicates the presence of significant amount of the dark matter.

The objects 1 and 2 are on stage of merging of two dwarf galaxies. 
It is indicated by structure of distribution of HI cloud and high difference in the velocity in the clumps in the galaxy.

The dwarfs in the group form the chain of the galaxies. 
We see the group in the process of the formation. 
Low metallicity of the gas in the galaxies support idea of "youth" of the galaxies. 
In addition we have found another candidate for very low-metallicity galaxy.
The chain shape of galaxies indicates that the group had not yet virialized. Thus we see young emerging group of dwarf galaxies.

\begin{acknowledgement}
This work was supported by the RFBR grants 08--02--00627 and 10--02--09373.
\end{acknowledgement}

\end{document}